\documentclass[aps,showpacs,prb,floatfix,twocolumn]{revtex4}
\usepackage{amsmath,amssymb,graphicx,bm,epsfig,color}

\begin{document}

\title{A new family of iron pnictides: BaFeAs$_2$ and BaFeSb$_2$}
\author{J. H. Shim, K. Haule, and G. Kotliar}

\affiliation{Department of Physics, Rutgers University, Piscataway, 
             NJ 08854, USA}

\begin{abstract}
  We investigate the structural, electronic, and magnetic properties
  of the hypothetical compound BaFe$Pn_2$ ($Pn$ = As and Sb), which is
  isostructural to the parent compound of the high temperature
  superconductor LaFeAsO$_{1-x}$F$_x$. Using density functional
  theory, we show that the Fermi surface, electronic structure and the
  spin density wave instability of BaFe$Pn_2$ are very similar to the
  Fe based superconductors.  Additionally, there are very dispersive
  metallic bands of a spacer $Pn$ layer, which are almost decoupled
  from Fe$Pn$ layer.  Our results show that experimental study of
  BaFe$Pn_2$ can test the role of charge and polarization fluctuation
  as well as the importance of two dimensionality in the mechanism of 
  superconductivity.
\end{abstract}
\date{\today}
\maketitle

The discovery of superconductivity in iron oxypnictide
LaFeAsO$_{1-x}$F$_x$~\cite{kamihara08} has generated renewed interest
in the phenomena of high temperature superconductivity, and brought
the transition metal based compounds to the forefront of condensed
matter research. The basic units responsible for the superconductivity
are the fluorite type [Fe$_2Pn_2$] layers, with $Pn$ a pnictide
element (P, As, Sb, and Bi). These layers are separated by spacer
layers, which play the role of charge reservoir. In the fluorite type
layers, the Fe atoms are surrounded by four pnictide atoms, forming a
tetrahedron.  The first class of materials studied has the ZrCuSiAs
structure (1111 compounds), where the spacer layer [$Ln_2$O$_2$] has
the "antifluoride" or Pb$_2$O$_2$ structure.  With $Ln$=Sm, critical
temperature higher than 55~K has been achieved\cite{sm}.

Superconductivity with $T_c = 38$~K was also found in the ternary
compounds $A$Fe$_2$As$_2$\cite{ba122} (122 compounds) with
ThCr$_2$Si$_2$ structure.  In this structure, the spacer layer is
provided by an alkali earth element $A$ = Ca, Sr, or Ba. Doping is
accomplished by substitution of $A$ by an alkali metal such as K or
Cs.

Superconductivity with $T_c = 18$~K was also achieved in
LiFeAs\cite{lifeas}.  The crystal structure of this compound is of the
PbFCl type, and consists on parallel Fe$_2$As$_2$ layers, separated by
a Li double layer spacer.

Iron based superconductivity has also been realized in binary iron
chalcogenides.  In the $\alpha$-FeSe$_{1-x}$ compound under pressure,
transition temperature can reach $27$~K~\cite{fese}.  In this compound
the need for charge reservoir spacer layer is eliminated, and
deviation from the Fe $d^6$ configuration is obtained by Se vacancies.
The iron layers are very similar to the ones found in the iron
pnictide materials.

Band structure calculations for these compounds revealed fairly
quasi-two-dimensional  bands near the Fermi level with a large $d$
character and a characteristic Fermi surface featuring two
electron pockets around the M point and two hole pockets around
the $\Gamma$ point in the Brillouin zone~\cite{singh1}.

\begin{figure}[tb]
\includegraphics[width=1.0\linewidth]{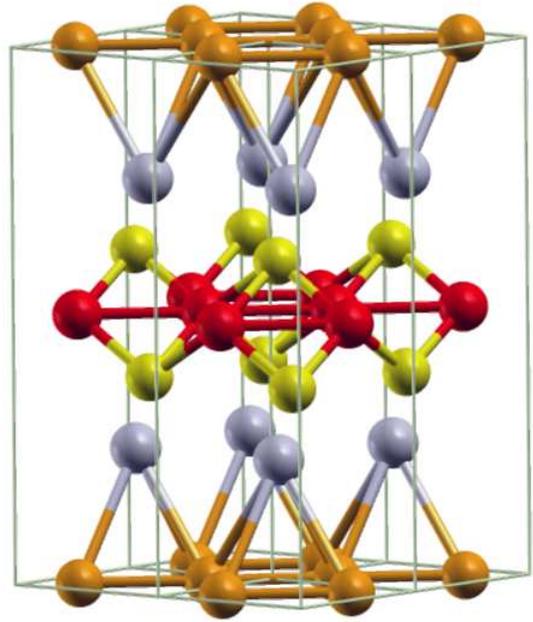}
\caption{
Crystal structure of BaFe$Pn_2$ ($Pn$ = As and Sb). Ba, Fe, $Pn$(2) and $Pn$(2)
atoms are denoted by gray, red, orange and yellow spheres, respectively.
There are two type of $Pn$; $Pn$(1) in B$Pn$ layer and $Pn$(2) in Fe$Pn$
layer. Isostructural compound LaFeAsO has the similar structure
of LaO layer with Ba$Pn$(1) layer.
}
\label{structure}
\end{figure}

In all these examples, the spacer layer is insulating (or absent in
the binary iron chalcogenides). The bands derived from the spacer
layers are far from the Fermi level. Thus the role of the spacer layer
is to provide a charge reservoir which keeps the charge of the FeAs
unit near -1. In an ionic picture, this corresponds to As$^{3-}$ and
Fe$^{2+}$ i.e. the Fe atoms are in a $d^6$ configuration.

There is no concensus yet on the mechanism of superconductivity in
this system. Our early calculation \cite{haule1}, as well as the more
extensive studies of Refs.~\onlinecite{no_phonon}, rule out a phonon
mechanism. However, both spin and orbital fluctuations
\cite{Mazin,RPA1,Elihu}, as well as the polarizability of the As
\cite{sawatzky}, have been suggested as potential candidates for the
pairing glue.  Understanding the mechanism of superconductivity and
the factors controlling the critical temperature is of great
fundamental interest, as well as of practical importance. Hence,
numerous related compounds are being explored theoretically and
experimentally. It is therefore important to correlate theoretical
quantities (i.e. density of states and bandwidth) and structural
information (i.e.  $c/a$ ratio, volume, and As-Fe-As bond angle) with
experimentally measured $T_c$.

In this letter, we identify two hypothetical compounds (112 compounds)
in the iron pnictide class: BaFeAs$_2$ and BaFeSb$_2$. They are
isostructural with LaFeAsO, as shown in Fig.~\ref{structure}.
Although they have not yet been synthesized, their isostructural
compounds $LnM$As$_2$ and $LnM$Sb$_2$ have been studied intensively
with various kind of rare earth $Ln$ and transition metals
$M$\cite{brylak95,wollesen96,zeng97}.  So, one can expect that BaFeSb$_2$ as
well as BaFeAs$_2$ can be synthesized.  We will show here, using
density functional theory (DFT) calculation, that the fermiology
derived from FeAs layers, is very similar to those of previously
studied iron pnictides. There is however, a very important difference
between all known Fe superconductors and these two compounds. Namely
the spacer layer Ba$Pn$ is {\it metallic}.

This is of interest because it allows to modify the interaction
strength in the charge channel without altering significantly the one
body part of the Hamiltonian. Its experimental investigation will shed
further light on the importance of correlations in the iron pnictides
compounds and on the role of charge and polarization fluctuations in
the pairing.

In order to predict the structural, electronic, and magnetic
structures of hypothetical compound BaFe$_2$$Pn_2$ ($Pn$ = As, Sb), we
used the full-potential linearized augmented planwave (FP-LAPW) method
as implemented in WIEN2k code\cite{wien}.  The generalized gradient
approximation (GGA) has been used for the exchange correlation
function\cite{gga}.  1000-$k$ points are used for the full Brillouin
zone integration.  Employed muffin-tin radii for Ba, Fe, As, and Sb
are 2.5, 2.24, 1.98, and 2.09 a.u., respectively.  The lattice
parameters as well as internal positions have been optimized by the
total energy minimization assuming spin density wave (SDW) 
ground state with tetragonal structure.
It was known in Ref.~\onlinecite{Yildirim} that this magnetic ground state
leades to very precise structural parameters. The
calculated structural parameters are shown in Table I and compared to
the experimental structural parameters of LaFeAsO.

\begin{table}
  \centering
  \caption{Calculated structural parameters of BaFe$Pn_2$ ($Pn$=As and Sb).
The space group is $P4/nmm$ with the internal coordinates of 
Ba (0.25, 0.25, $z_{Ba}$), Fe (0.75, 0.25, 0.5), $Pn$(1) (0.75, 0.25, 0.0),
and $Pn$(2) (0.25, 0.25, $z_{Pn}$). For comparison, 
the structural parameters of LaFeAsO is extracted from Ref.~\onlinecite{kamihara08}.}
  \label{occupancy}
  \begin{tabular}{cccccccc}
    \hline\hline
    ~~~ & BaFeAs$_2$& BaFeSb$_2$& LaFeAsO\\
    \hline
    $a$ (\AA) & 3.999 &  4.430 & 4.035 \\
    $c$ (\AA) & 11.826 & 11.977 & 8.741 \\
    $z_{Ba}$,$z_{La}$ & 0.2299 & 0.2395 & 0.142 \\
    $z_{Pn}$ & 0.6129 & 0.6177 & 0.651 \\
    Fe-$Pn$(2) length (\AA) & 2.404 & 2.625 & 2.412 \\
    Ba-$Pn$(1) length (\AA) & 3.384 & 3.624 & 2.367 \\
    $Pn$(2)-Fe-$Pn$(2) angle ($^\circ$)& 112.54 & 115.05 & 113.55 \\
    \hline\hline
  \end{tabular}
\end{table}

There is a strong similarity between the BaFeAs$_2$ and
LaFeAsO$_{1-x}$F$_x$ structure.
%
%The structure of FeAs layer shows strong similarity between BaFeAs$_2$
%and LaFeAsO$_{1-x}$F$_x$.  
%
The Fe-As bond length and angle in BaFeAs$_2$ are 2.404\AA~ and
112.54$^\circ$, which are comparable to the values 2.412\AA~ and
113.55$^\circ$ in LaFeAsO$_{1-x}$F$_x$.
%
%We notice here that the structural parameters for the new compounds
%were obtained by the LDA optimization in the paramagnetic phase, which
%might not be very precise due to deficiency of LDA, as notice in
%Ref.~\cite{Yildirim}. 
%Indeed the LDA structural optimization for LaFeAsO
%gives angle of 120.25$^\circ$, rather than 113.55$^\circ$.
%
%This reflects the local crystal structure of FeAs layer and the 
%valence of Fe and As atoms are not affected by the change of other layer.
However, there is a clear difference between BaAs layers and LaO
layers in LaFeAsO$_{1-x}$F$_x$, although the structural group is the
same.  The Ba-As(1) bond length is much larger than the La-O bond
length, such that As(1) atoms form a single layer separated from Ba
atoms.  This increases the FeAs interlayer distance, and the lattice
constant along $c$-direction becomes 35 \% larger than the value of
LaFeAsO$_{1-x}$F$_x$. This indicates that BaFeAs$_2$ might have more
two-dimensional properties than other Fe based superconductors.

Substitution of As by Sb significantly modifies the structure.  Fe-Sb
bond length increases to 2.625 \AA~ due to the large size of Sb ion,
which substantially increases the lattice parameter $a$ of BaFeSb$_2$.
%This will give a 
%clear change of band dispersion as will be discussed in the following.
The Sb-Fe-Sb bond angle in BaFeSb$_2$ is 115.05$^\circ$ and is larger
than corresponding angle in BaFeAs$_2$. It has been suggested that the
bond anlge correlates with the transition temperature of Fe based
superconductors\cite{kreyssig08}. The highest $T_c$ seems to be
achieved in the structures with the tetrahderon angle closest to the
ideal tetrahderon angle of 109.47$^\circ$. This would suggest that
doped BaFeSb$_2$ might have lower $T_c$ than doped LaFeAsO and BaFeAs$_2$.

\begin{figure}[tb]
\includegraphics[angle=270,width=1.0\linewidth]{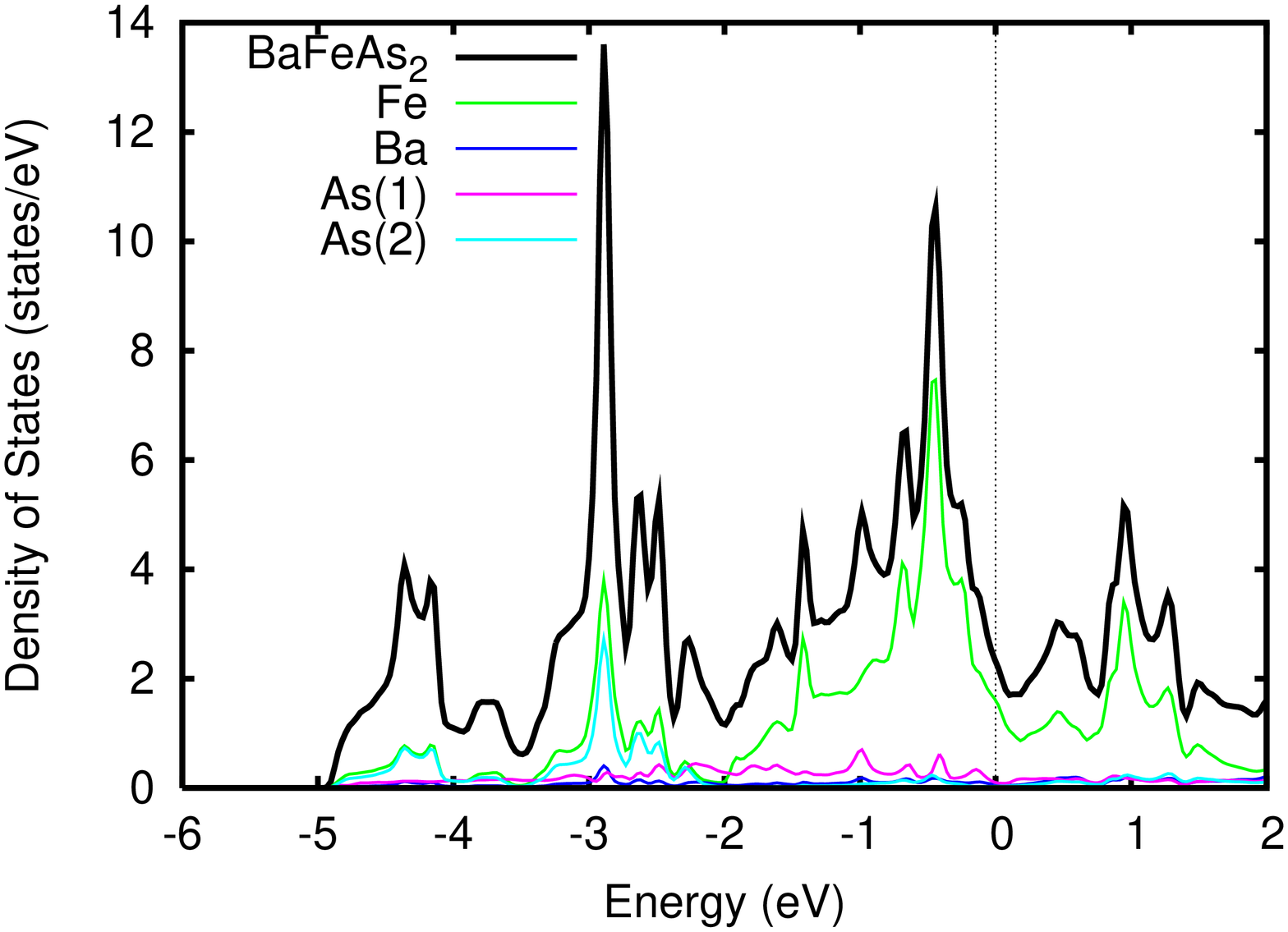}
\includegraphics[angle=270,width=1.0\linewidth]{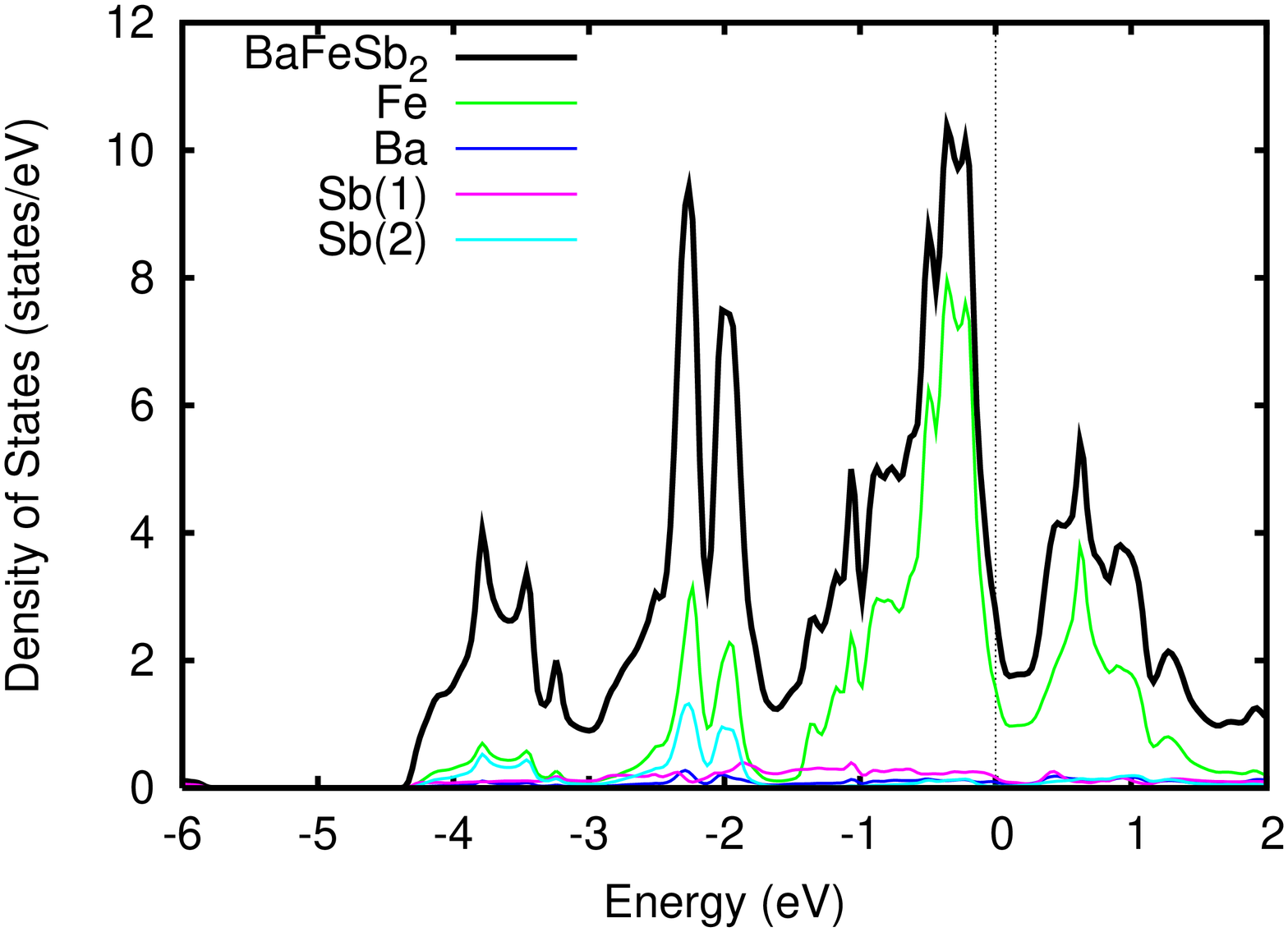}
\caption{
Total and partial DOS of BaFeAs$_2$ and BaFeSb$_2$.
}
\label{dos}
\end{figure}

In order to the check electronic structure of BaFe$Pn_2$, we perform
the calculation of density of states (DOS), band structures, and Fermi
surfaces in paramagnetic phase.  Figure~\ref{dos} shows total and partial 
DOS of BaFeAs$_2$ and BaFeSb$_2$. The general shape of Fe and
As partial DOS in FeAs layer show clear similarity between BaFeAs$_2$
and LaFeAsO.  The DOS between -2 eV and 2 eV is dominated by Fe $3d$
states, and Sb $4p$ states are located around -3 eV, and their
contribution at the Fermi level is negligible. At the Fermi level, the
total DOS shows the same negative slope with the minimum slightly
above the Fermi level, similar to the other FeAs superconductors
within DFT.

The electronic structure of BaAs spacer layer shows a clear difference
compared to LaO layer in LaFeAsO. The latter shows a clear insulating
gap at the Fermi level, while the former is metallic. Due to the large
density of Fe-d states at $E_F$, the spacer layer however contributes
small percentage of states at $E_F$.
Because the large distance between the As(1) and the FeAs layer, the
hybridization between As(1) and FeAs layer is negligible.  In order to
check the role of As(1) atoms, we calculated the electronic structure
of BaFeAs in the absence of As(1) atoms with the same crystal
structure of BaFeAs$_2$. The shape of partial Fe DOS does not change
with the absence of As(1), which means that the As(1) atoms behave as
separated single layer. Their role is hence to increase the spacing
between the FeAs layers in z-direction.

As expected from the structural parameters, the electronic structure
of BaFeSb$_2$ is distinctly different from BaFeAs$_2$.  Due to larger
in-plane distance of Fe atoms, the width of Fe $3d$ partial DOS is
substantially reduced and is confined between -1.5 eV and 1 eV.  Hence
the Fe moment in the magnetic state is expected to be large in
BaFeSb$_2$, and the electronic correlations more important. Indeed the
moment of Fe in the most stable magnetic state is  $2.61\,\mu_B$
compared to $2.12\,\mu_B$ in BaFeAs$_2$.  Hence the Fe-d electron are
more localized in the structure with larger pnictide $Pn$.

%\begin{figure}[tb]
%\includegraphics[angle=270,width=1.0\linewidth]{dos_as_3d.ps}
%\includegraphics[angle=270,width=1.0\linewidth]{dos_sb_3d.ps}
%\caption{
%$3d$ partial DOS of BaFeAs$_2$ and BaFeSb$_2$
%}
%\label{dos2}
%\end{figure}

\begin{figure}[tb]
\includegraphics[width=1.0\linewidth]{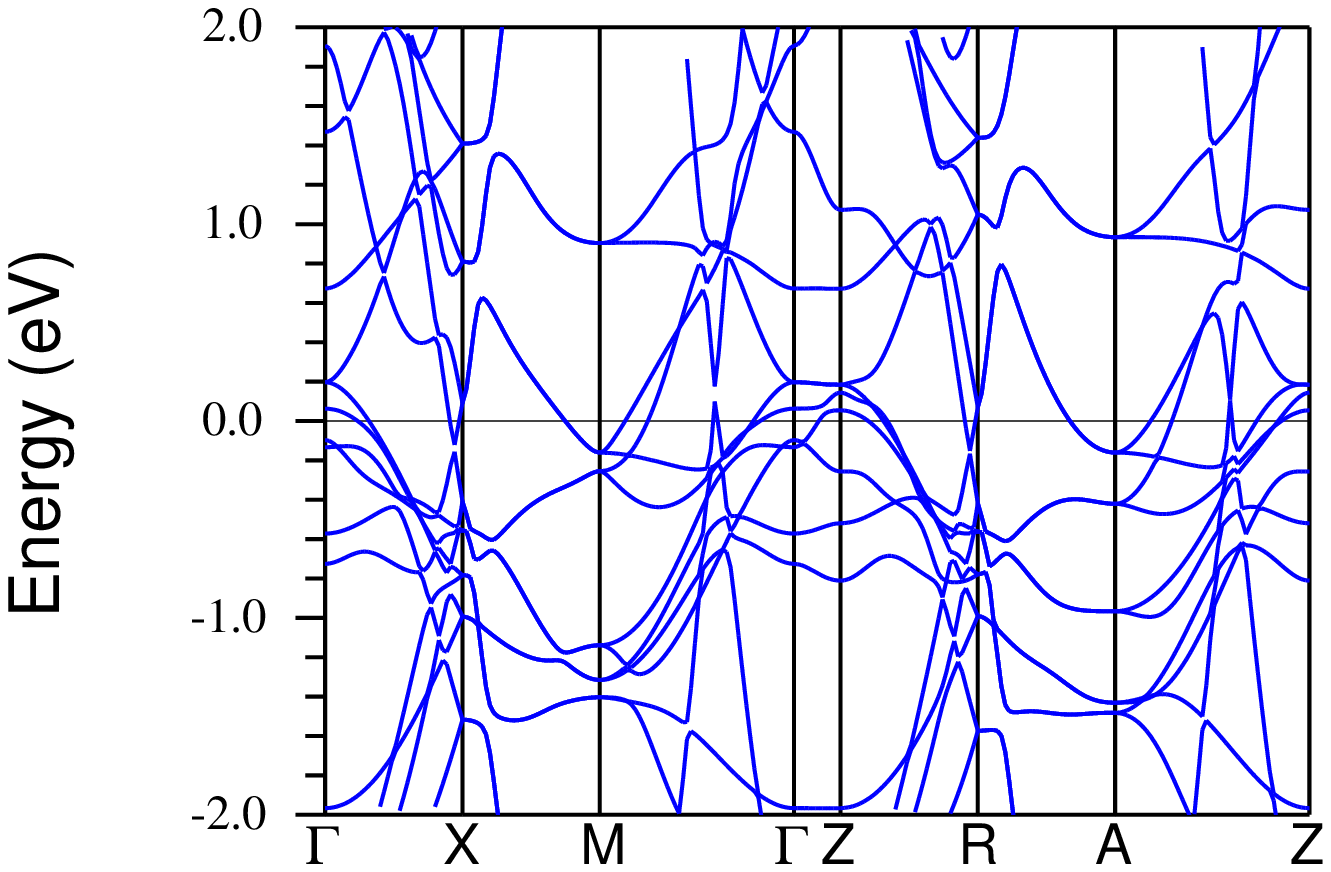}
\includegraphics[width=1.0\linewidth]{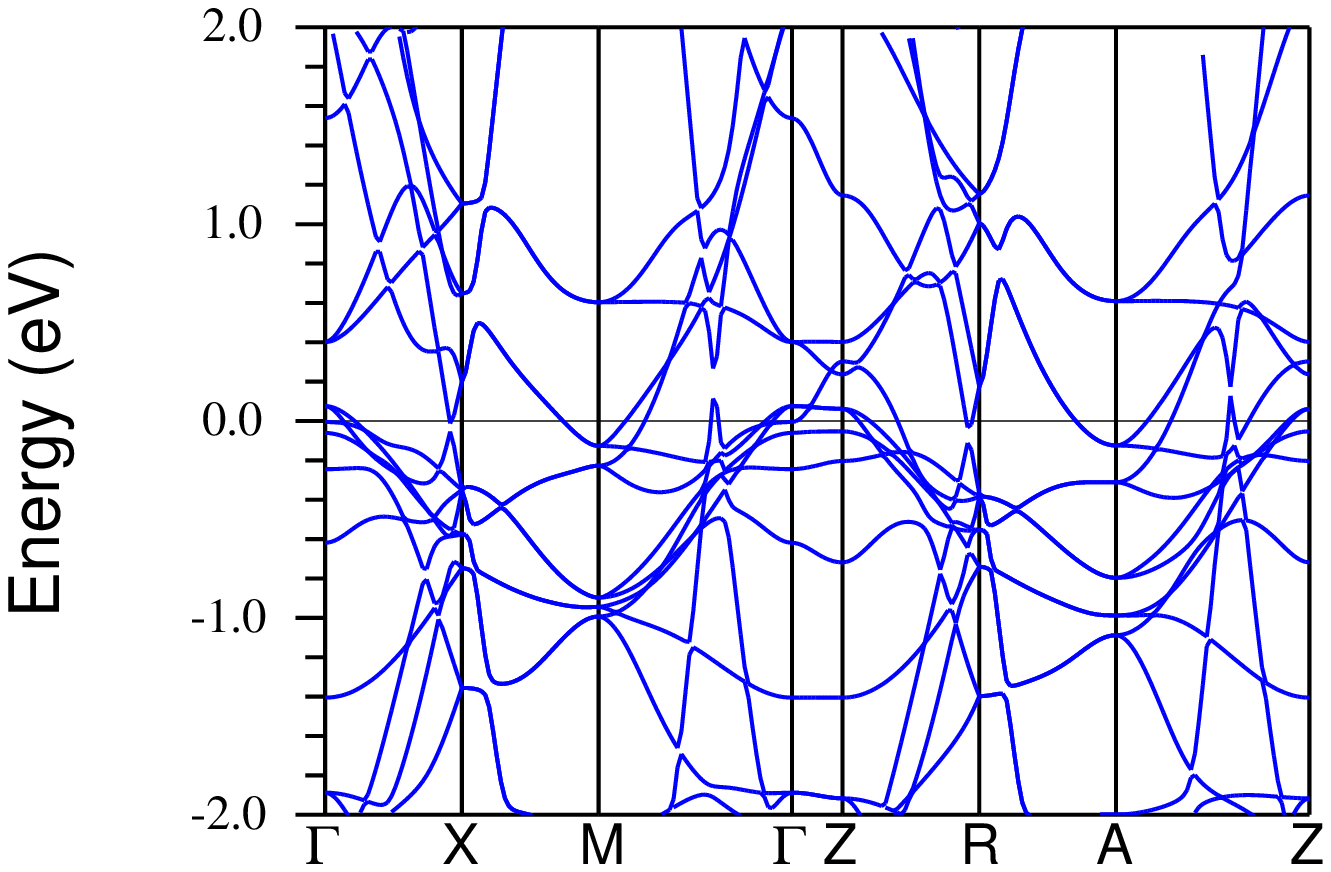}
\caption{
Band structure of BaFeAs$_2$ and BaFeSb$_2$
}
\label{band}
\end{figure}

The band structures of BaFeAs$_2$ and BaFeSb$_2$ in shown in
Fig.~\ref{band}. It shows many common feature observed in other Fe
based superconductors.  There are two hole pockets at $\Gamma$ point
and two electron pockets at M point. 
% I REALLY DO NOT SEE THAT
%The dispersionless bands along
%$\Gamma$-M symmetry lines indicate the quasi-two-dimensionality of
%these compounds.
%
There is however a distinctive difference between the band structure
of other known Fe superconductors and the two compounds studied here.
There are a few very dispersive bands crossing the Fermi level, which
are primarily comming from the As(1) atoms in the spacer layer. Near
the Fermi level, these bands are however well decoupled from the bands
comming from the FeAs layer.

%\begin{figure}[tb]
%\includegraphics[width=1.0\linewidth]{fband_as.ps}
%\includegraphics[width=1.0\linewidth]{fband_sb.ps}
%\caption{
%Fe $3d$ fat band structure of BaFeAs$_2$ and BaFeSb$_2$
%}
%\label{band2}
%\end{figure}

\begin{figure}[tb]
\includegraphics[width=0.8\linewidth]{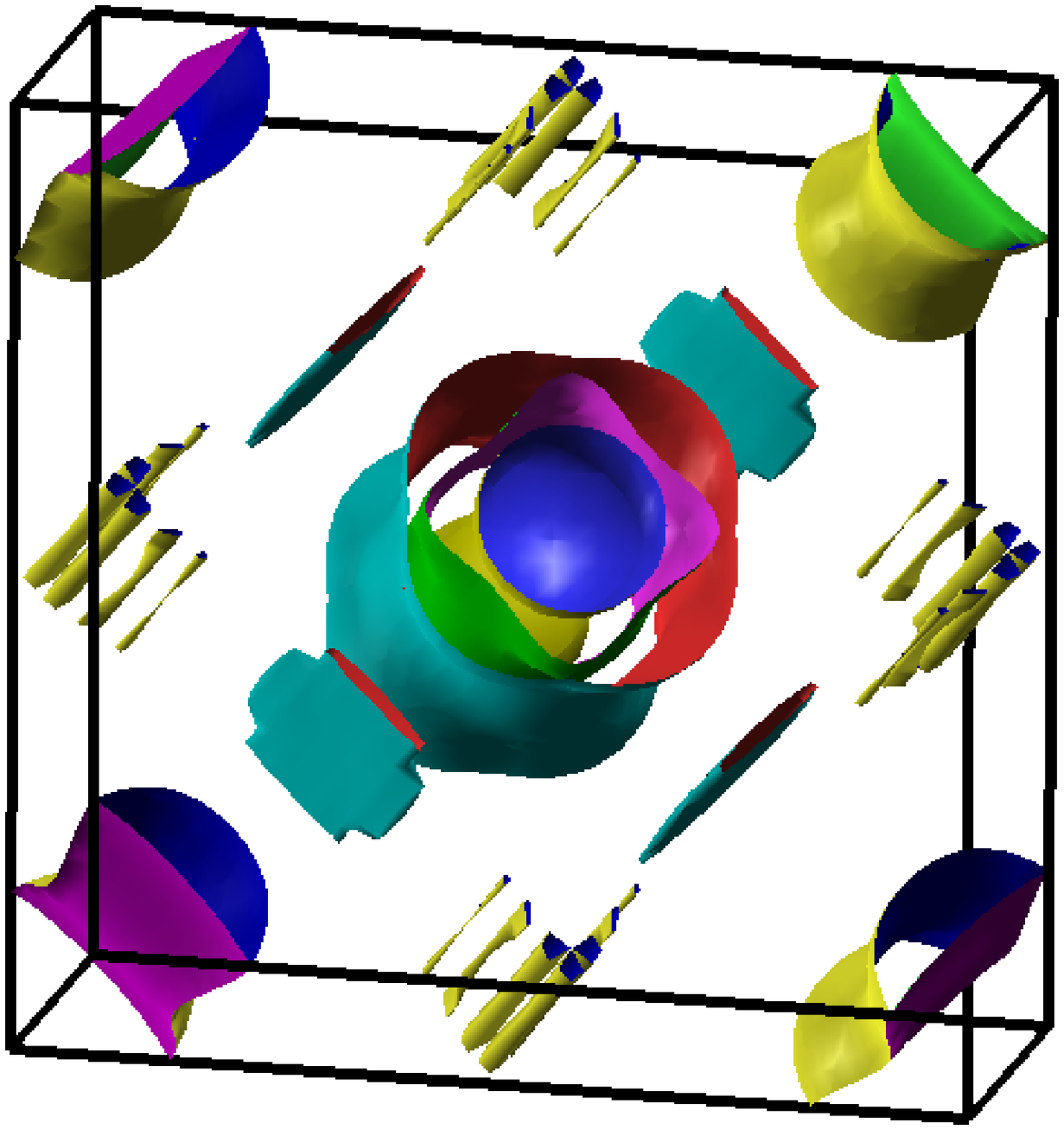}
\includegraphics[width=0.8\linewidth]{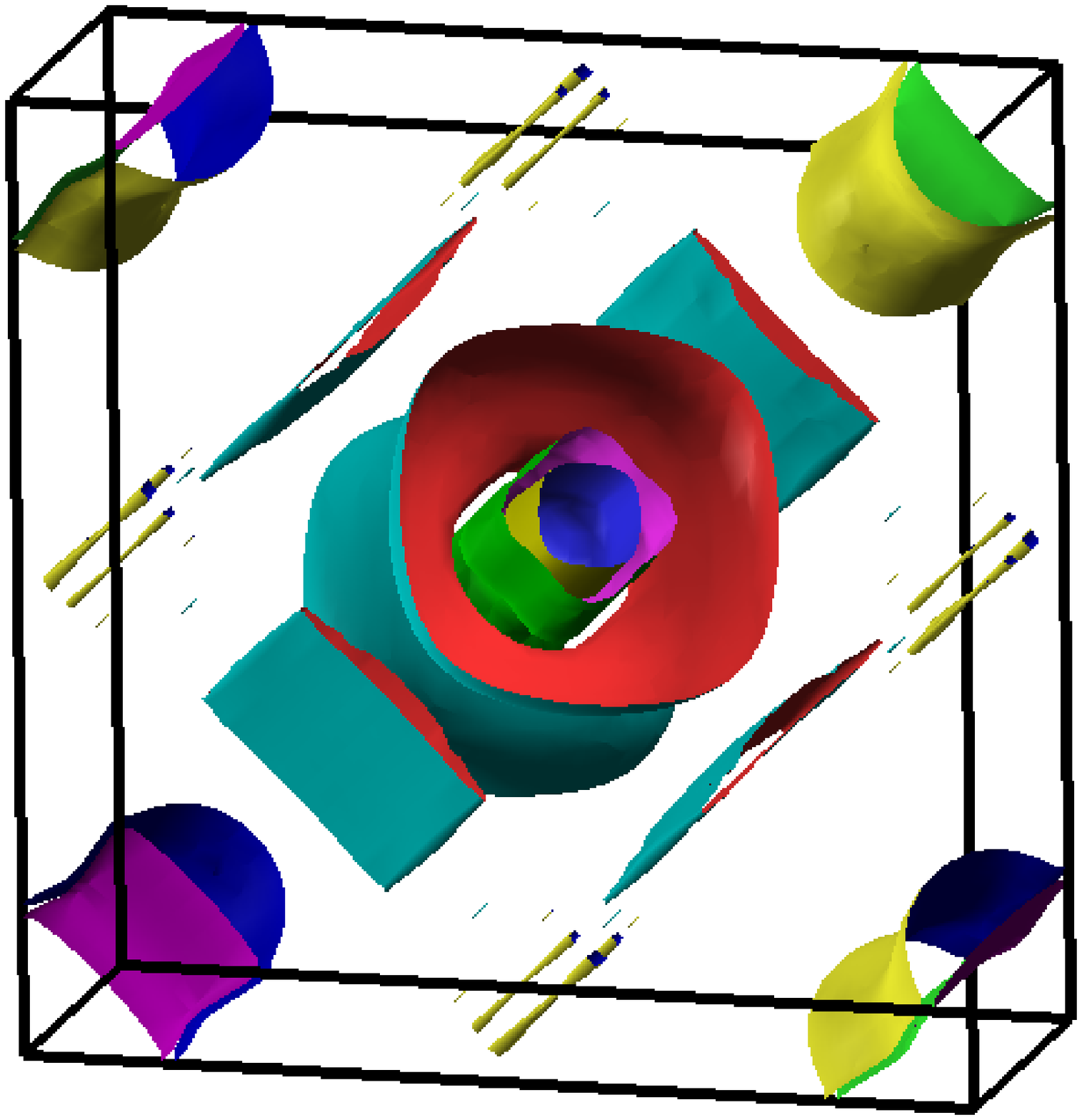}
\caption{
Fermi surfaces of BeFeAs$_2$ and BaFeSb$_2$ in the first Brillouin zone
centered at the $\Gamma$ point.
}
\label{fs}
\end{figure}

Due to the similarity of band structure, the Fermi surface of
BaFeAs$_2$ also shows clear similarity to that of Fe based
superconductors.  Figure~\ref{fs}(a) shows the quasi two-dimensional
Fermi surfaces with cyliderical shape.  There are two hole pockets at
$\Gamma$ points and two electron pockets at M point with the high
degree of nesting between the hole and electron pockets.  
% I DO NOT UNDERSTAND THAT
% Even the
%intersecting Fermi surface structures around M point is same feature
%to the case of other Fe based superconductors.  
Due to the As(1) states at the Fermi level, there are additional Fermi
surfaces with diamond shape connecting the X points. They clearly have
a very two dimensional character.

%This Fermi surface come from the dispersive As bands, also with
%two-dimensional behavior.
%In BaFeSb$_2$, these Fermi surfaces are merged with hole pocket around
%$\Gamma$ point.

% The role of additional Fermi surfaces coming from BaAs layer should be
% investigated further. Because it is two-dimensional and clear nesting
% feature along ($\pi$, $\pi$, 0) direction, there can be some
% instability along this direction. This might give a charge density
% wave in the As(1) layer and a lower structural symmetry than $P4/nmm$.
% However, we excluded this possibility in this study because other
% $LnM$As$_2$ and $LnM$Sb$_2$ have been observed in stable tetragonal
% symmetry with the space group of $P4/nmm$.

%??????
Although the mechanism for Fe based superconductivity has not yet been
established, the superconductivity is believed to be associated with
the magnetism. All parent compounds of Fe based superconductors show
first order structural transition from tetragonal to orthorombic
phase, as well as the SDW transition at somewhat
lower temperature. With doping or application of pressure, the two
transitions are suppressed and the superconductivity emerges.

We checked the stability of the magnetic states in the two compounds.
We found that stripe-type SDW phase is the most stable magnetic
configuration among the comensurate magnetic states of a double unit
cell.  
The stabilization energy of the SDW phase compared to the paramegnetic
phase in BaFeAs$_2$ is $E_{SDW}$=-161 meV/Fe and in BaFeSb$_2$ is -434
meV/Fe. The magnetic moments in the muffin tin radius of Fe atom is
2.12 $\mu_B$ in case of BaFeAs$_2$ and 2.61 $\mu_B$ in BaFeSb$_2$.
The SDW magnetism is enhaced by the enhanced in-plane lattice
constant.  

% Also the magnetism is very sensitive to the internal atomic
% position of As(2), which is already discussed in literature.

The LDA+DMFT calculations have demonstrated that while the FeAs
compounds are less correlated than the cuprates, they are not too far
from the localization-delocalization transition \cite{haule1}. As a
result, it was found that at high temperatures LaFeAsO has a bad
semimetal incoherent regime. We expect this regime to extend to higher
dopings and pressures in the BaFeSb$_2$ compound, where the ratio of
Hund's coupling and the crystal fields splitting is larger than in
LaFeAsO or BaFeAs$_2$.

% [ THIS REALLY DOES NOT FIT YET
% 
% It has been argued on the basis of LDA+DMFT calculations, that even
% though the Fermi surfaces can be reproduced by tight binding fits
% involving a smaller number of bands, the local density of states
% within an electron volt window of the Fermi level involves all the
% five 3d bands~\cite{haule1}. Hence a five band model describing the Fe
% $3d$ orbitals is the minimal model needed to describe the FeAs layers.
% % I don't think this is good! They found superconductivity in the two
% % band model as well :  http://arxiv.org/pdf/0805.3343
% % but they have different symmetry than in
% % http://xxx.lanl.gov/pdf/0807.0498v3
% 
% %This observation is consistent with renormalization group studies,
% %starting from the weak coupling end, which account for
% %superconductivity only when five bands are included in the
% %calculation~\cite{renormalization}.
% 
% We suggested that in the parent compounds, the correlations are weaker
% than the cuprates, in the sense that the interaction parameters
% (Hubbard $U$ and Hunds rule $J$) place this material on the itinerant
% side of the Mott transition. Still correlation effects are
% significant, which account for significant amount of incoherent
% optical spectral weight in the optical conductivity, renormalization
% of the band structure, and high resistivities characteristic of a bad
% metal. The bad metallic regime is well reproduced by parameters $U$ =
% 4 eV and $J$ = 0.3 eV\cite{haule2}.
% ]
% 

It is generally believed that two dimensionality is beneficial to
superconductivity \cite{lonzarich}.  This compound, being more two
dimensional than previously considered ones, if one uses the distance
between the FeAs layers as a criteria, would then be a good candidate
for high temperature superconductivity. On the other hand, the
superconducting temperature in 1111 compounds clearly decreases with
decreasing $c/a$ ratio, which would then suggest a lower $T_c$ for the
new 112 family of compounds.
%
%
%On the other hand, it has been noticed that the critical temperature
%of the 1111 and the 122 family decreases as the Fe-As-Fe angle
%deviates from the ideal tetrahedral coordination \cite{kreyssig08}, 
%which would then suggest a lower $T_c$ for the 112 family.
%
Synthesis of memebers of the 112 family, would help elucidate further
the role of two dimensionality in this compounds.

To summarize, we have suggested the synthesis of a new family of
compounds, the 112 BaFe$Pn_2$, with $Pn$ a pnictide. Doping could be
achieved, by substitution of Ba by K. Using electronic structure
tools, we have shown that these compounds have the same basic building
block as the other known iron pnictide superconductors.  We
identified, however, a key difference in the character of the spacer
block, which can provide new hints as to the mechanism of
superconductivity and the basic low energy physics of this compounds.

The spin fluctuations of these compounds are very similar to the other
iron based superconductors,
so we expect very similar variations of $T_c$ with doping and
pressure if a spin flucuation mechanism is operational. On the other
hand, in a framework where the superconductivity is mediated by
electronic polarizability, the presence of additional metallic layers,
will strongly modify the superconducting transition temperature.

We predict that the magnetic moment is larger in the Sb member of the
family which is \textit{less} nested than its As relative.
Experimental confirmation of this prediction would rule out weak
coupling approaches to this problem. Experimental determination of the
$T_c$, in these two compounds would clarify if nesting or the size of
the magnetic moment is a better predictor of the superconducting
transition temperature.  Given the frantic pace of the research in
this field we trust that the experimental answers to these important
questions will be known in the very near future.
 
Acknowledgment: We acknowledge useful discussions with S. W. Cheong.
This work was supported by the NSF division of materials research.

\end{document}